\input harvmac

\def\hat{\widehat}
\def\CF{{\cal F}}
%
\let\includefigures=\iftrue
%
%
%
\newfam\black
\input rotate
\input epsf
\noblackbox
%
%
\includefigures
\message{If you do not have epsf.tex (to include figures),}
\message{change the option at the top of the tex file.}
\def\figin{\epsfcheck\figin}\def\figins{\epsfcheck\figins}
\def\epsfcheck{\ifx\epsfbox\UnDeFiNeD
\message{(NO epsf.tex, FIGURES WILL BE IGNORED)}
\gdef\figin##1{\vskip2in}\gdef\figins##1{\hskip.5in}
\else\message{(FIGURES WILL BE INCLUDED)}%
\gdef\figin##1{##1}\gdef\figins##1{##1}\fi}
\def\DefWarn#1{}

\def\figinsert{\goodbreak\midinsert}
\def\ifig#1#2#3{\DefWarn#1\xdef#1{fig.~\the\figno}
\writedef{#1\leftbracket fig.\noexpand~\the\figno}%
\figinsert\figin{\centerline{#3}}\medskip\centerline{\vbox{\baselineskip12pt
\advance\hsize by -1truein\noindent\footnotefont{\bf
Fig.~\the\figno:} #2}}
\bigskip\endinsert\global\advance\figno by1}
\else
\def\ifig#1#2#3{\xdef#1{fig.~\the\figno}
\writedef{#1\leftbracket fig.\noexpand~\the\figno}%
\global\advance\figno by1} \fi
\def\yboxit#1#2{\vbox{\hrule height #1 \hbox{\vrule width #1
\vbox{#2}\vrule width #1 }\hrule height #1 }}
\def\fillbox#1{\hbox to #1{\vbox to #1{\vfil}\hfil}}
\def\ybox{{\lower 1.3pt \yboxit{0.4pt}{\fillbox{8pt}}\hskip-0.2pt}}

\def\rightarrowbox#1#2{
  \setbox1=\hbox{\kern#1{${ #2}$}\kern#1}
  \,\vbox{\offinterlineskip\hbox to\wd1{\hfil\copy1\hfil}
    \kern 3pt\hbox to\wd1{\rightarrowfill}}}

\def\half{{1\over 2}}
\def\Tr{{{\rm Tr~ }}}
\def\tr{{\rm tr\ }}

\def\CF{{\cal F}}

\def\CN{{\cal N}}
\def\CO{{\cal O}}

\def\CS{{\cal S}}

\def\tilde{\widetilde}

\def\II{\relax{I\kern-.10em I}}

\def\IZ{\relax\ifmmode\mathchoice
{\hbox{\cmss Z\kern-.4em Z}}{\hbox{\cmss Z\kern-.4em Z}}
{\lower.9pt\hbox{\cmsss Z\kern-.4em Z}} {\lower1.2pt\hbox{\cmsss
Z\kern-.4em Z}}\else{\cmss Z\kern-.4em Z}\fi}
\def\IB{\relax{\rm I\kern-.18em B}}
\def\IC{{\relax\hbox{$\inbar\kern-.3em{\rm C}$}}}
\def\ID{\relax{\rm I\kern-.18em D}}
\def\IE{\relax{\rm I\kern-.18em E}}
\def\IF{\relax{\rm I\kern-.18em F}}
\def\IG{\relax\hbox{$\inbar\kern-.3em{\rm G}$}}
\def\IGa{\relax\hbox{${\rm I}\kern-.18em\Gamma$}}
\def\IH{\relax{\rm I\kern-.18em H}}
\def\II{\relax{\rm I\kern-.18em I}}
\def\IK{\relax{\rm I\kern-.18em K}}
\def\IN{\relax{\rm I\kern-.18em N}}
\def\IP{\relax{\rm I\kern-.18em P}}

%
\def\inbar{\,\vrule height1.5ex width.4pt depth0pt}

\font\cmss=cmss10 \font\cmsss=cmss10 at 7pt
\def\IR{\relax{\rm I\kern-.18em R}}

\def\lp10{l_P^{10}}
\def\lp11{l_P^{11}}
\def\R11{R_{11}}

%
%
\lref\VenezianoAH{ G.~Veneziano and S.~Yankielowicz, ``An
Effective Lagrangian For The Pure N=1 Supersymmetric Yang-Mills
Theory,'' Phys.\ Lett.\ B {\bf 113}, 231 (1982).
}

\lref\GopakumarKI{ R.~Gopakumar and C.~Vafa, ``On the gauge
theory/geometry correspondence,'' Adv.\ Theor.\ Math.\ Phys.\
{\bf 3}, 1415 (1999) [arXiv:hep-th/9811131].
}

\lref\VafaWI{ C.~Vafa, ``Superstrings and topological strings at
large N,'' J.\ Math.\ Phys.\  {\bf 42}, 2798 (2001)
[arXiv:hep-th/0008142].
}

\lref\HofmanBI{ C.~Hofman, ``Super Yang-Mills with flavors from
large N(f) matrix models,'' arXiv:hep-th/0212095.
}

\lref\BenaTN{ I.~Bena, S.~de Haro and R.~Roiban, ``Generalized
Yukawa couplings and matrix models,'' arXiv:hep-th/0212083.
}

\lref\OhtaRD{ K.~Ohta, ``Exact mesonic vacua from matrix models,''
arXiv:hep-th/0212025.
}

\lref\OokouchiBE{ Y.~Ookouchi, ``N = 1 gauge theory with flavor
from fluxes,'' arXiv:hep-th/0211287.
}

\lref\FengZB{ B.~Feng, ``Seiberg duality in matrix model,''
arXiv:hep-th/0211202.
}

\lref\BenaUA{ I.~Bena, R.~Roiban and R.~Tatar, ``Baryons,
boundaries and matrix models,'' arXiv:hep-th/0211271.
}

\lref\NaculichHR{ S.~G.~Naculich, H.~J.~Schnitzer and N.~Wyllard,
``Matrix model approach to the N = 2 U(N) gauge theory with
matter in the  fundamental representation,'' arXiv:hep-th/0211254.
}

\lref\ArgurioHK{ R.~Argurio, V.~L.~Campos, G.~Ferretti and
R.~Heise, ``Baryonic corrections to superpotentials from
perturbation theory,'' arXiv:hep-th/0211249.
}

\lref\FengYF{ B.~Feng and Y.~H.~He, ``Seiberg duality in matrix
models. II,'' arXiv:hep-th/0211234.
}

\lref\TachikawaWK{ Y.~Tachikawa, ``Derivation of the Konishi
anomaly relation from Dijkgraaf-Vafa with  (bi-)fundamental
matters,'' arXiv:hep-th/0211189.
}

\lref\DemasureSC{ Y.~Demasure and R.~A.~Janik, ``Effective matter
superpotentials from Wishart random matrices,''
arXiv:hep-th/0211082.
}

\lref\BenaKW{ I.~Bena and R.~Roiban, ``Exact superpotentials in N
= 1 theories with flavor and their matrix  model formulation,''
arXiv:hep-th/0211075.
}

\lref\SuzukiGP{ H.~Suzuki, ``Perturbative derivation of exact
superpotential for meson fields from  matrix theories with one
flavour,'' arXiv:hep-th/0211052.
}

\lref\McGreevyYG{ J.~McGreevy, ``Adding flavor to
Dijkgraaf-Vafa,'' arXiv:hep-th/0211009.
}

\lref\ArgurioXV{ R.~Argurio, V.~L.~Campos, G.~Ferretti and
R.~Heise, ``Exact superpotentials for theories with flavors via a
matrix integral,'' arXiv:hep-th/0210291.
}

\lref\BerensteinSN{ D.~Berenstein, ``Quantum moduli spaces from
matrix models,'' arXiv:hep-th/0210183.
}

\lref\HollowoodZK{ T.~J.~Hollowood and T.~Kingaby, ``The phase
structure of mass-deformed SU(2) x SU(2) quiver theory,''
arXiv:hep-th/0210096.
}

\lref\agmoo{ O.~Aharony, S.~S.~Gubser, J.~M.~Maldacena, H.~Ooguri
and Y.~Oz, ``Large N field theories, string theory and gravity,''
Phys.\ Rept.\  {\bf 323}, 183 (2000) arXiv:hep-th/9905111.}

\lref\NovikovEE{V.~A.~Novikov, M.~A.~Shifman, A.~I.~Vainshtein
and V.~I.~Zakharov, ``Instanton Effects In Supersymmetric
Theories,'' Nucl.\ Phys.\ B {\bf 229}, 407 (1983).
}

\lref\ArgyresJJ{ P.~C.~Argyres and M.~R.~Douglas, ``New phenomena
in SU(3) supersymmetric gauge theory,'' Nucl.\ Phys.\ B {\bf
448}, 93 (1995) [arXiv:hep-th/9505062].
}

\lref\ArgyresXN{ P.~C.~Argyres, M.~Ronen Plesser, N.~Seiberg and
E.~Witten, ``New N=2 Superconformal Field Theories in Four
Dimensions,'' Nucl.\ Phys.\ B {\bf 461}, 71 (1996)
[arXiv:hep-th/9511154].
}

\lref\DijkgraafXD{ R.~Dijkgraaf, M.~T.~Grisaru, C.~S.~Lam,
C.~Vafa and D.~Zanon, ``Perturbative computation of glueball
superpotentials,'' arXiv:hep-th/0211017.
}

\lref\SeibergRS{ N.~Seiberg and E.~Witten, ``Electric - magnetic
duality, monopole condensation, and confinement in N=2
supersymmetric Yang-Mills theory,'' Nucl.\ Phys.\ B {\bf 426}, 19
(1994) [Erratum-ibid.\ B {\bf 430}, 485 (1994)]
[arXiv:hep-th/9407087].
}

\lref\kovner{A. Kovner and M. Shifman, ``Chirally Symmetric Phase
Of Supersymmetric Gluodynamics,'' Phys. Rev. {\bf D56} (1997)
2396, hep-th/9702174.}

\lref\konishione{ K.~Konishi, ``Anomalous Supersymmetry
Transformation Of Some Composite Operators In Sqcd,'' Phys.\
Lett.\ B {\bf 135}, 439 (1984).
}

\lref\konishitwo{ K.~i.~Konishi and K.~i.~Shizuya, ``Functional
Integral Approach To Chiral Anomalies In Supersymmetric Gauge
Theories,'' Nuovo Cim.\ A {\bf 90}, 111 (1985).
}

\lref\arkmur{N. Arkani-Hamed and H. Murayama, hep-th/9707133.}

\lref\bipz{ E.~Brezin, C.~Itzykson, G.~Parisi and J.~B.~Zuber,
``Planar Diagrams,'' Commun.\ Math.\ Phys.\  {\bf 59}, 35 (1978).
}

\lref\BershadskyCX{ M.~Bershadsky, S.~Cecotti, H.~Ooguri and
C.~Vafa, ``Kodaira-Spencer theory of gravity and exact results
for quantum string amplitudes,'' Commun.\ Math.\ Phys.\  {\bf
165}, 311 (1994) [arXiv:hep-th/9309140].
}

\lref\mmreview{ P.~Ginsparg and G.~W.~Moore, ``Lectures On 2-D
Gravity And 2-D String Theory,'' arXiv:hep-th/9304011.
}

\lref\gorsky{ A.~Gorsky, ``Konishi anomaly and N = 1 effective
superpotentials from matrix models,'' arXiv:hep-th/0210281.
}

\lref\CachazoJY{ F.~Cachazo, K.~A.~Intriligator and C.~Vafa, ``A
large N duality via a geometric transition,'' Nucl.\ Phys.\ B {\bf
603}, 3 (2001) [arXiv:hep-th/0103067].
}

\lref\KutasovVE{ D.~Kutasov, ``A Comment on duality in N=1
supersymmetric nonAbelian gauge theories,'' Phys.\ Lett.\ B {\bf
351}, 230 (1995) [arXiv:hep-th/9503086].
}

\lref\FerrariJP{ F.~Ferrari, ``On exact superpotentials in
confining vacua,'' arXiv:hep-th/0210135.
}

\lref\WittenXI{ E.~Witten, ``The Verlinde Algebra And The
Cohomology Of The Grassmannian,'' arXiv:hep-th/9312104, and in
{\it Quantum Fields And Strings: A Course For Mathematicians},
ed. P. Deligne et. al. (American Mathematical Society, 1999),
vol. 2, pp. 1338-9.
}

\lref\SeibergPQ{ N.~Seiberg, ``Electric - magnetic duality in
supersymmetric nonAbelian gauge theories,'' Nucl.\ Phys.\ B {\bf
435}, 129 (1995) [arXiv:hep-th/9411149].
}

\lref\IntriligatorID{ K.~A.~Intriligator and N.~Seiberg,
``Duality, monopoles, dyons, confinement and oblique confinement
in supersymmetric SO(N(c)) gauge theories,'' Nucl.\ Phys.\ B {\bf
444}, 125 (1995) [arXiv:hep-th/9503179].
}

\lref\FujiWD{ H.~Fuji and Y.~Ookouchi, ``Comments on effective
superpotentials via matrix models,'' arXiv:hep-th/0210148.
}

\lref\KutasovVE{ D.~Kutasov, ``A Comment on duality in N=1
supersymmetric nonAbelian gauge theories,'' Phys.\ Lett.\ B {\bf
351}, 230 (1995) [arXiv:hep-th/9503086].
}

\lref\KutasovNP{ D.~Kutasov and A.~Schwimmer, ``On duality in
supersymmetric Yang-Mills theory,'' Phys.\ Lett.\ B {\bf 354},
315 (1995) [arXiv:hep-th/9505004].
}

\lref\KutasovSS{ D.~Kutasov, A.~Schwimmer and N.~Seiberg,
``Chiral Rings, Singularity Theory and Electric-Magnetic
Duality,'' Nucl.\ Phys.\ B {\bf 459}, 455 (1996)
[arXiv:hep-th/9510222].
}

\lref\IntriligatorAU{ K.~A.~Intriligator and N.~Seiberg,
``Lectures on supersymmetric gauge theories and
electric-magnetic  duality,'' Nucl.\ Phys.\ Proc.\ Suppl.\  {\bf
45BC}, 1 (1996) [arXiv:hep-th/9509066].
}

\lref\DijkgraafFC{ R.~Dijkgraaf and C.~Vafa, ``Matrix models,
topological strings, and supersymmetric gauge theories,''
arXiv:hep-th/0206255.
}

\lref\DijkgraafVW{ R.~Dijkgraaf and C.~Vafa, ``On geometry and
matrix models,'' arXiv:hep-th/0207106.
}

\lref\DijkgraafDH{ R.~Dijkgraaf and C.~Vafa, ``A perturbative
window into non-perturbative physics,'' arXiv:hep-th/0208048.
}

\lref\IntriligatorJR{ K.~A.~Intriligator, R.~G.~Leigh and
N.~Seiberg, ``Exact superpotentials in four-dimensions,'' Phys.\
Rev.\ D {\bf 50}, 1092 (1994) [arXiv:hep-th/9403198].
}

\lref\SeibergBZ{ N.~Seiberg, ``Exact results on the space of
vacua of four-dimensional SUSY gauge theories,'' Phys.\ Rev.\ D
{\bf 49}, 6857 (1994) [arXiv:hep-th/9402044].
}

\lref\CDSW{ F.~Cachazo, M.~R.~Douglas, N.~Seiberg and E.~Witten,
``Chiral rings and anomalies in supersymmetric gauge theory,''
arXiv:hep-th/0211170.
}

\lref\DijkgraafPP{ R.~Dijkgraaf, S.~Gukov, V.~A.~Kazakov and
C.~Vafa, ``Perturbative analysis of gauged matrix models,''
arXiv:hep-th/0210238.
}

\lref\GopakumarWX{ R.~Gopakumar, ``${\cal N}=1$ Theories and a
Geometric Master Field,'' arXiv:hep-th/0211100.
}

\lref\Schnitzer{S.G.~ Naculich, H.J.~ Schnitzer and N.~Wyllard,
``The $\CN=2$ $U(N)$ gauge theory prepotential and periods from a
perturbative matrix model calculation,'' arXiv:hep-th/0211123.}

\lref\SeibergVC{ N.~Seiberg, ``Naturalness versus supersymmetric
nonrenormalization theorems,'' Phys.\ Lett.\ B {\bf 318}, 469
(1993) [arXiv:hep-ph/9309335].
}

\lref\DouglasNW{ M.~R.~Douglas and S.~H.~Shenker, ``Dynamics of
SU(N) supersymmetric gauge theory,'' Nucl.\ Phys.\ B {\bf 447},
271 (1995) [arXiv:hep-th/9503163].
}

\lref\CachazoPR{ F.~Cachazo and C.~Vafa, ``N = 1 and N = 2
geometry from fluxes,'' arXiv:hep-th/0206017.
}

\lref\IntriligatorJR{ K.~A.~Intriligator, R.~G.~Leigh and
N.~Seiberg, ``Exact superpotentials in four-dimensions,'' Phys.\
Rev.\ D {\bf 50}, 1092 (1994) [arXiv:hep-th/9403198].
}

\lref\DijkgraafXD{ R.~Dijkgraaf, M.~T.~Grisaru, C.~S.~Lam,
C.~Vafa and D.~Zanon, ``Perturbative Computation of Glueball
Superpotentials,'' arXiv:hep-th/0211017.
}
\lref\superspace{ S.~J.~Gates, M.~T.~Grisaru, M.~Rocek and
W.~Siegel, ``Superspace, Or One Thousand And One Lessons In
Supersymmetry,'' Front.\ Phys.\  {\bf 58}, 1 (1983)
[arXiv:hep-th/0108200].
}

\lref\nicolai{ H.~Nicolai, ``On A New Characterization Of Scalar
Supersymmetric Theories,'' Phys.\ Lett.\ B {\bf 89}, 341 (1980).
}

\lref\migdal{ A.~A.~Migdal, ``Loop Equations And 1/N Expansion,''
Phys.\ Rept.\  {\bf 102}, 199 (1983).
}

\lref\staudacher{ M.~Staudacher, ``Combinatorial solution of the
two matrix model,'' Phys.\ Lett.\ B {\bf 305}, 332 (1993)
[arXiv:hep-th/9301038].
}

\lref\voiculescu{{\it Free Probability Theory}, ed. D. Voiculescu,
pp. 21--40, AMS, 1997.}

\lref\CeresoleZS{ A.~Ceresole, G.~Dall'Agata, R.~D'Auria and
S.~Ferrara, ``Spectrum of type IIB supergravity on AdS(5) x
T(11): Predictions on N  = 1 SCFT's,'' Phys.\ Rev.\ D {\bf 61},
066001 (2000) [arXiv:hep-th/9905226].
}

\lref\ofer{O.~Aharony, unpublished.}

\newbox\tmpbox\setbox\tmpbox\hbox{\abstractfont 
}
\Title{\vbox{\baselineskip12pt\hbox to\wd\tmpbox{\hss
hep-th/0212225}
}} {\vbox{\centerline{Adding Fundamental Matter to }
\smallskip
\centerline{``Chiral Rings and Anomalies }
\centerline{in Supersymmetric Gauge Theory''}}}
\smallskip
\centerline{Nathan Seiberg}
\smallskip
\bigskip
\centerline{School of Natural Sciences, Institute for Advanced
Study, Princeton NJ 08540 USA}
\bigskip
\vskip 1cm
 \noindent
We consider a supersymmetric $U(N)$ gauge theory with matter
fields in the adjoint, fundamental and anti-fundamental
representations.  As in the framework which was put forward by
Dijkgraaf and Vafa, this theory can be described by a matrix
model.  We analyze this theory along the lines of [F.~Cachazo,
M.~Douglas, N.S. and E.~Witten, ``Chiral Rings and Anomalies in
Supersymmetric Gauge Theory'' hep-th/0211170] and show the
equivalence of the gauge theory and the matrix model. In
particular, the anomaly equations in the gauge theory is
identified with the loop equations in the matrix model.

\Date{December 2002}
%

%
%

\newsec{Introduction}

Recently Dijkgraaf and Vafa \DijkgraafDH\ were motivated by
earlier work
\refs{\BershadskyCX\GopakumarKI\VafaWI\CachazoJY\CachazoPR
\DijkgraafFC-\DijkgraafVW} to conjecture an interesting relation
between SUSY gauge theories and matrix models.   Many authors
have added matter in the fundamental representation
\refs{\HollowoodZK\BerensteinSN
\ArgurioXV\McGreevyYG\SuzukiGP\BenaKW\DemasureSC
\TachikawaWK\ArgurioHK\NaculichHR\BenaUA
\FengZB\FengYF\OokouchiBE\OhtaRD\BenaTN-\HofmanBI} to this
framework. We will examine a general theory of this form which
includes the various examples of \refs{\HollowoodZK-\HofmanBI} as
special cases.  We will follow the point of view of \CDSW, and
will extend it to this case with fundamental matter.

We consider an ${\cal N}=1$ supersymmetric $U(N)$ gauge theory
with matter in the adjoint $\Phi$, $N_f$ fundamentals $Q^f$ and
$N_f$ anti-fundamentals $\tilde Q_{\tilde f}$ ($f$ and $\tilde f$
are the flavor indices). The tree level superpotential is
 \eqn\treesupa{W_{tree}=\Tr W(\Phi) + \tilde Q_{\tilde f}
 m^{\tilde f}_f (\Phi) Q^f}
where we suppressed the color indices.  The function $W$ and the
matrix $m$ are taken to be polynomials
 \eqn\coeffmg{\eqalign{
 W(z)&=\sum_{k=1}^n {1\over k+1} g_k z^{k+1} \cr
 m_f^{\tilde f}(z) &=\sum_{k=1}^{l+1}  m_{f, k}^{\tilde f}z^{k-1}}}

We consider arbitrary $N$ and $N_f$.   For $nN_f >N$ these
theories exhibit a certain duality
\refs{\KutasovVE\KutasovNP-\KutasovSS} exchanging $N
\leftrightarrow nN_f - N$, and we hope that this discussion will
shed light on it.

In section 2 we analyze this quantum field theory focusing on its
chiral ring, the anomaly equations and the low energy effective
superpotential.  In section 3 we consider the corresponding
matrix model and prove its equivalence to the gauge theory.

\newsec{Gauge theory considerations}

\subsec{Anomaly equations}

As in \CDSW\ we will be interested in the chiral operators
 \eqn\chiralopf{\eqalign{
 T(z)&= \Tr {1\over z-\Phi} \cr
 w_\alpha(z)&= {1 \over 4\pi} \Tr {W_\alpha\over z-\Phi} \cr
 R(z)&= -{1 \over 32 \pi^2} \Tr {W_\alpha W^\alpha \over
 z-\Phi}\cr
 M_{\tilde f}^f(z) &= \tilde Q_{\tilde f} {1\over z-\Phi} Q^f}}
Since $W_\alpha Q^f$ and $\tilde Q_{\tilde f} W_\alpha$ are not
in the chiral ring, there is no need to include more
operators\foot{We thank E.~Witten for a useful discussion on this
point.}.  Also, since the gauge group is $U(N)$ rather than
$SU(N)$ we do not include ``baryonic operators.''

We perform the following five independent transformations in the
functional integral
 \eqn\transf{\eqalign{
 \delta \Phi&= {1\over z-\Phi} \cr
 \delta \Phi&= {W_\alpha \over z-\Phi} \cr
 \delta \Phi&= {W_\alpha W^\alpha \over z-\Phi} \cr
 \delta Q^f &= {1\over z-\Phi} \lambda^f_{f'} Q^{f'}\cr
 \delta \tilde Q_{\tilde f} &=  \tilde  \lambda^{\tilde
 f'}_{\tilde f}\tilde Q_{\tilde f'}{1\over z-\Phi} \cr}}
where $\lambda^f_{f'}$ and $ \tilde \lambda^{\tilde f'}_{\tilde
f}$ are constant flavor matrices.  The anomaly equations of these
transformations are
 \eqn\anoef{\eqalign{
 &\Tr {W'(\Phi) \over z-\Phi} +\tilde Q_{\tilde f}
 {m'^{\tilde f}_f (\Phi)  \over z-\Phi }Q^f = 2R(z)T(z)
 +w_\alpha(z) w^\alpha(z)\cr
 &{1 \over 4\pi} \Tr {W'(\Phi) W_\alpha \over z-\Phi}  =
 2R(z)w_\alpha(z)\cr
 &-{1\over 32 \pi^2}\Tr {W'(\Phi) W_\alpha W^\alpha \over z-\Phi}
 = R(z)^2\cr
 & \lambda^f_{f'} \tilde Q_{\tilde f}
 {m^{\tilde f}_f (\Phi)  \over z-\Phi }Q^{f'} =\lambda^f_{f}
 R(z)\cr
 & \tilde \lambda^{\tilde f'}_{\tilde f} \tilde Q_{\tilde f'}
 {m^{\tilde f}_f (\Phi)  \over z-\Phi }Q^{f} = \tilde
 \lambda^{\tilde f}_{\tilde f} R(z)\cr
 }}
where the equalities are only in the chiral ring; i.e.\ they are
up to SUSY commutators.  The $\CO(1/ z)$ term in the first
equation is the equation of motion of $\Tr \Phi$.  The $\CO(1/
z^2)$ term in that equation and the $\CO(1/ z)$ terms in the
fourth and fifth equations are the Konishi anomaly
\refs{\konishione,\konishitwo} for $\Phi$, $Q$ and $\tilde Q$
respectively.  The other equations and the other powers of $1/z$
are various generalizations.

Since $W(\Phi)$ and $m_f^{\tilde f}(\Phi)$ are polynomials, we
can replace these equations with
 \eqn\anoeff{\eqalign{
 &\left[W'(z)T(z)\right]_- + \tr \left[m'(z)M(z) \right]_- =
 2R(z)T(z) +w_\alpha(z) w^\alpha(z)\cr
 &\left[W'(z) w_\alpha(z)\right]_-  = 2R(z)w_\alpha(z)\cr
 &\left[W'(z) R(z)\right]_-  = R(z)^2\cr
 & \left[\big(M(z)m (z)\big)_f^{f'} \right ]_-= R(z)
 \delta_f^{f'}  \cr
 & \left[\big(m(z) M(z)\big)_{\tilde f}^{\tilde f'} \right]_-
 = R(z) \delta_{\tilde f}^{\tilde f'} \cr
 }}
Here $m(z)$ and $M(z)$ are matrices in flavor space.  We multiply
such matrices as $(AB)^f_{f''}= A^f_{f'} B^{f'}_{f''}$ and $\tr$
denotes a trace over the flavor indices. In the last two
equations we used the fact that the corresponding equations in
\anoef\ are satisfied for every $\lambda$ and $\tilde \lambda$.
These identities are also true for the expectation values of the
operators \chiralopf.

Writing the third equation in \anoeff\ as
 \eqn\anoeffthi{W'(z) R(z) +{1 \over 4} f(z)  = R(z)^2}
with a polynomial $f(z)=-4\left[W'(z) R(z)\right]_+$ of degree
$n-1$, we solve for $R(z)$
 \eqn\solveR{2R(z)= W'(z) -\sqrt{W'(z)^2 +f(z)}}
Similarly, we write the last two equations in \anoeff\ as
  \eqn\anoefffi{\eqalign{
 & \big( M(z)m (z) -q(z)\big)_f^{f'}=  R(z)\delta_f^{f'} \cr
 & \big(m(z) M(z) -\tilde q(z)\big)_{\tilde f}^{\tilde f'} =
  R(z)\delta_{\tilde f}^{\tilde f'} \cr
 }}
in terms of polynomial matrices
 \eqn\qtqdef{\eqalign{
 q(z)&=\left[M(z) m(z)\right]_+ \cr
 \tilde q(z)&=\left[m(z) M(z)\right]_+ }}
We can now solve for the matrix $M(z)$
 \eqn\anoai{M(z)=   R(z)m^{-1}(z) +q(z) m^{-1}(z) =
 R(z)m^{-1}(z) -m^{-1}(z) \tilde q(z) }
where $m^{-1}(z)$ is the inverse matrix.  For consistency the
polynomial matrices $q(z)$ and $\tilde q(z)$ are related by
$\tilde q=mqm^{-1}$.  Using the solution for $R(z)$ in \solveR\
the expression for $M(z)$ \anoai\ leads to a solution for $M(z)$
in terms of the polynomials $f(z)$ and $q(z)$.

The polynomial $f(z)$ is kept arbitrary, but the polynomials
$q(z)$ are fixed as follows.  From \anoai\ it is clear that $M(z)$
is singular when $m(z)$ has a zero eigenvalue.  The freedom in
the polynomials $q(z)$ is exactly such as to remove these
singularities.  To see this, consider for simplicity the case
$N_f=1$, where the flavor matrices are one dimensional. $m(z)$ is
a polynomial of degree $l$ \coeffmg, and therefore it has $l$
zeroes $z_I$.  For large $z$ the resolvent $M(z)$ behaves as
$1\over z$ and therefore $q(z)=\left[M(z) m(z)\right]_+ $ is a
polynomial of degree $l-1$.  The $l$ coefficients of $q$ can be
tuned to set the $l$ residues of $M(z)$ at $z_I$ to zero.

Now that we have solved for $R(z)$ and $M(z)$, it is simple to
solve for $w_\alpha(z)$ and $T(z)$ using the first two equations
in \anoeff.  In doing that we need to introduce new polynomials
 \eqn\rhoc{\eqalign{\rho_\alpha(z) &=-4 [W'(z)w_\alpha(z)]_+ \cr
 c(z)&=-4[W'(z)T(z)]_+}}
Note that the solution for $R(z)$ is as in the theory without the
fundamental matter, but $T(z)$ is different; it depends on $m(z)$.

\subsec{The effective superpotential}

We parametrize the superpotential of the adjoint field as
 \eqn\superpotai{W'(z)= g_n\prod_{i=1}^n (z-a_i)}
in terms of its stationary points $a_i$. We consider the classical
vacuum
 \eqn\Phiex{\langle \Phi \rangle =\pmatrix{
 a_1&&&&&&&&&\cr
 &.  &&&&&&&&\cr
 &&.  &&&&&&&\cr
 &&&a_2&&&&&&\cr
 &&&&.  &&&&&\cr
 &&&&&.  &&&&\cr
 &&&&&&a_3&&&\cr
 &&&&&&&.  &&\cr
 &&&&&&&&.  &\cr
 &&&&&&&&&a_n\cr}}
where $a_i$ occurs $N_i$ times.  We assume that $W(\Phi)$ and
$m(\Phi)$ are generic and in particular $m(a_i)$ does not have a
zero eigenvalue.  Since all the matter fields acquire masses, the
low energy degrees of freedom are only the gauge fields in
$\prod_i U(N_i)$.  Following \CDSW\ we express them as the gauge
invariant objects
 \eqn\contin{\eqalign{
 S_i &= {1 \over 2\pi i} \oint_{C_i} R(z) dz\cr
 w_\alpha^{(i)}&= {1 \over 2\pi i}\oint_{C_i} w_\alpha(z) dz\cr
 N_i&= {1 \over 2\pi i} \oint_{C_i} T(z) dz}}
where $C_i$ is a contour around $a_i$.

Since we assumed that $m$ is generic, all the quarks are massive
and are integrated out.  Therefore, our effective Lagrangian will
not depend on the ``meson operator in the $i$ group''
 \eqn\Midef{ M^{(i)}= {1 \over 2\pi i} \oint_{C_i} M(z) dz}

As in \CDSW\ (see also \DijkgraafXD) we integrate out the massive
fields in perturbation theory.  It is straightforward to add the
fundamental matter to the discussion of the Feynman diagrams in
\CDSW. Assume for simplicity that the group $U(N)$ is unbroken;
i.e.\ $N_1=N$. Then considerations of symmetry and holomorphy
along the lines of \SeibergVC\ constrain the perturbative
effective superpotential to be of the form
 \eqn\effsupfo{W_{eff}^{pert}= W_\alpha^2 F\left({g_k
 W_{\alpha}^{k-1} \over g_1^{k+1\over 2}}, {m_{f,k}^{\tilde f}
 W_{\alpha}^{k-1} \over (m_{f,k=1}^{\tilde f})^{k+1\over
 2}}\right)}
Since the theory depends on several masses $g_1$,
$m_{f,k=1}^{\tilde f}$, the function $F$ can also depend on
various ratios of them.  From \effsupfo\ we see that the number of
$W_\alpha$ in a diagram that contributes to $W_{eff}^{pert}$ is
constrained to be
 \eqn\numberwal{\# W_\alpha = 2 + \sum_r ( k_r-1)}
Here $ r$ labels the vertices of the Feynman diagram, and $k_r$ is
the index $k$ either for $g_k$ or for $m_{f,k}^{\tilde f}$ at the
vertex. We now write the diagram using 'tHooft's index loops
notation. Every $\Phi$ has a double line and every $Q$ or $\tilde
Q$ has a single line. The number of index loops in the diagram is
 \eqn\numindl{L=2 + {1\over 2} \sum_r ( k_r-1) -2h -b}
where $h$ is the number of handles in the surface and $b$ is the
number of boundaries ($b$ is the number of $Q$ or $\tilde Q$
index loops). From \numberwal\ and \numindl\
 \eqn\numberwalf{\# W_\alpha= 2L -2 + 4h +2b}
As in \CDSW, it is straightforward to extend this discussion to
the case where the gauge group is broken, and the external light
fields in the diagram are $W_\alpha^{(i)}$. Here we find
 \eqn\numberwalf{\sum_i\# W^{(i)}_\alpha= 2L -2 + 4h +2b}
We also know that chiral operators have at most two factors of
$W_\alpha^{(i)}$ in a trace and therefore there are at most two
$W_\alpha^{(i)}$ on each index loop:
 \eqn\ineqwl{\sum_i\# W^{(i)}_\alpha \le 2L}
Combining \numberwalf\ and \ineqwl\ we learn that only three kinds
of diagrams can contribute to the effective superpotential:
\item{1.} $h=b=0$.  These diagrams have the topology of a sphere.
One index loop has no operator insertion (leading to a factor of
$N_i$) and every other index loop has an insertion of $S_j$
leading to a result proportional to $N_iS_j^{L-1}$ ($S_j^{L-1}$
stands for a product of $L-1$ factors of $S_j$ not necessarily all
with the same value of $j$).
\item{2.} $h=b=0$. These diagrams also have the topology of a
sphere. Two index loops have an insertion of
$w_\alpha^{(i)}$ on each, and every other index loop has $S_j$
leading to a result proportional to $w^{(i)}_\alpha w^{(k)\alpha }
S_j^{L-2}$.
\item{3.} $h=0$, $b=1$.  These diagrams have the topology of a
disk.  They have an insertion of $S_j$ on each index loop leading
to a result proportional to $S_j^L$.

The sphere diagrams do not involve $Q$ and $\tilde Q$. Therefore,
their contribution is the same as in the theory without these
fields; i.e.\ having only the adjoint field $\Phi$.  This theory
has a shift symmetry associated with the decoupled diagonal
$U(1)$ field $\sum_i w_\alpha^{(i)}$ \CDSW. This symmetry is best
implemented by adding the auxiliary spinor coordinate
$\psi_\alpha$, and combining $S_i$, $w^{(i)}_\alpha$ and $N_i$ to
a superfield
 \eqn\calss{\CS_i(\psi)=S_i + \psi^\alpha w_\alpha^{(i)}
 -\psi^1\psi^2 N_i}
Then the symmetry of shifting $\psi^\alpha$ determines the sphere
contribution in terms of a single function $\CF_0^{pert}$ of
$\CS_i (\psi)$
 \eqn\sphereco{\int d^2 \psi \CF^{pert}_0 (\CS_i(\psi))}
The integral over $\psi$ leads to the two sphere contributions we
mentioned above. The function $\CF^{pert}_0$ is exactly the same
as this function in the theory without the fundamental matter. It
is given by
 \eqn\cfzpert{\CF^{pert}_0= \sum_i\half \CS_i^2 \log
 \left({g_n\prod _{j\not= i}( a_i- a_j)\over \Lambda_0}\right)
 - \sum_{i,j \atop i\not=j}\CS_i\CS_j \log {a_i-a_j \over
 \Lambda_0}+ \CO(S_j^3)}
Here we wrote explicitly the one loop result which depends on the
UV cutoff $\Lambda_0$ . The first term arises from integrating out
the massive chiral superfield in the adjoint of $U(N_i)$, and the
second term is from the massive vector superfields which acquired
mass in the Higgs mechanism.  The higher order terms $\CO(S_j^3)$
are independent of $\Lambda_0$.

The disk diagrams lead to another function of $S_i$ whose
contribution to the effective superpotential is not integrated
over $\psi$
 \eqn\cfopert{\CF^{pert}_1= \sum_i S_i \log \left(\det
 {m(a_i)\over \Lambda_0}\right) + \CO(S_j^2)}
where we wrote explicitly the one loop terms which arise from the
massive quarks $Q$ and $\tilde Q$.  Again, the higher order terms
$\CO(S_j^2)$ are independent of the UV cutoff $\Lambda_0$.
$\CF^{pert}_1$ is not integrated over $\psi$ and it does not
respect the shift symmetry of $\psi$.  This is consistent with
the fact that in the theory with fundamental matter the over all
$U(1)$ superfield $\sum_i w_\alpha^{(i)}$ does not decouple.

The full effective superpotential at the scale $\mu$ where the
gauge dynamics of the unbroken $\prod_i U(N_i)$ gauge fields is
still weak is
 \eqn\Weffp{W_{eff}^{pert}= \tau_0\sum_i S_i + \half \log \left(
 {\Lambda_0\over \mu}\right)^3 \sum_i \int d^2\psi \CS_i^2(\psi)
 +\int d^2 \psi \CF^{pert}_0(\CS_i(\psi)) + \CF^{pert}_1(S_i)}
The first term is the bare coupling.  The second term includes
the one loop running of the gauge couplings in $\prod_i SU(N_i)$;
it depends only on the $SU(N_i)$ fields $s_i=S_i + { 1\over 2N_i}
(w_\alpha^{(i)})^2$
 \eqn\bareaction{\half \log \left(
 {\Lambda _0 \over \mu}\right)^3 \sum_i \int d^2\psi\CS_i^2(\psi)
 = \sum_i N_i s_i \log \left({\Lambda_0 \over \mu}\right)^3}
The dependence on the UV cutoff $\Lambda_0$ from all the terms is
 \eqn\cutoffd{(2N-N_f)\log \Lambda_0 \sum_i S_i + \log \Lambda_0
 \left( \sum_i w_\alpha^{(i)}\right)^2}
In accordance with the one loop beta function we choose the bare
coupling
 \eqn\barec{\tau_0=-(2N-N_f)\log \Lambda_0/\Lambda }
with a finite scale $\Lambda$.  This has the effect of changing
$\Lambda_0 \to \Lambda$ everywhere except the term $\log
\Lambda_0 \left( \sum_i w_\alpha^{(i)}\right)^2$.  This indicates
that the over all $U(1)$ superfield $\sum_i w_\alpha^{(i)}$ is
free in the renormalized theory.  We will continue to use the
same notation as in \cfzpert-\bareaction\ but with $\Lambda_0\to
\Lambda$.

The strong IR dynamics is implemented by replacing \bareaction\
(after $\Lambda_0\to \Lambda$) with the Veneziano-Yankielowics
superpotential \VenezianoAH
 \eqn\VYS{W_{VY}(s_i)=\sum_i s_i\left(\log{\Lambda
 ^{3N_i} \over  s_i^{N_i}} +N_i\right)= \half \sum_i\int d^2
 \psi \CS_i^2\left(\log{\Lambda^3 \over \CS_i}+ {3\over 2}\right)}
Therefore, the final answer for the effective superpotential
is
 \eqn\weff{ W_{eff}(S_i,w_\alpha^{(i)}, N_i) =\int d^2
 \psi \CF_0(\CS_i(\psi)) + \CF_1(S_i) =\int d^2 \psi \left( \CF_0
 (\CS_i(\psi)) + \psi^2\psi^1 \CF_1(\CS_i(\psi)) \right)}
with
 \eqn\fullCFz{\eqalign{
 \CF_0&=\CF^{pert}_0 + \half \sum_i \CS_i^2\left(\log{
 \Lambda^3 \over \CS_i}+ {3\over 2}\right)\cr
 \CF_1&=\CF^{pert}_1}}
Here $s_i=S_i + { 1\over 2N_i} (w_\alpha^{(i)})^2$ are
independent chiral superfields, and $w_\alpha^{(i)}$ are
independent field strengths of the massless vector superfields.

Using the effective superpotential we can compute expectation
values of operators by differentiating with respect to the
coefficients in \coeffmg
 \eqn\phiexp{\eqalign{
 &{1\over k+1} \langle \Tr \Phi^{k+1} \rangle =
 {\partial W_{eff} \over \partial g_ k} ={\partial
  \over \partial g_k} \int d^2\psi \CF_0 + {\partial
 \CF_1 \over \partial g_k} \cr
 &\langle \tilde Q_{\tilde f} \Phi^{k-1}
  Q^f \rangle = {\partial W_{eff} \over \partial m_{f, k}^{\tilde
  f}} ={\partial \CF_1 \over \partial m_{f, k}^{\tilde f}}}}
In the second equation we used the fact that the sphere diagrams
do not involve the fundamental matter, and therefore the function
$\CF_0$ is independent of $ m_{f, k}^{\tilde f}$.  The sphere
contribution in the first equation is as in the theory without
the fundamental matter.  The disk contribution depends on $ m_{f,
k}^{\tilde f}$.  Therefore $\langle\Tr \Phi^{k+1} \rangle$ and
$\langle T(z) \rangle$ depend on $ m_{f, k}^{\tilde f}$ in accord
with the comment after \rhoc.

Since we have already solved for these expectation values using
the anomaly equations, we effectively found the full effective
superpotential.  The variables $S_i$ or more precisely $s_i$
capture the information in the coefficients in the polynomial
$f(z)$ in \anoeffthi.  Equivalently, the freedom in $f(z)$ is
determined by the values of the fields $s_i$.  The polynomials
$\rho_\alpha$ \rhoc\ are determined in terms of the variables
$w_\alpha^{(i)}$.  Finally, $c(z)$ \rhoc\ is determined in terms
of $N_i$.  We remarked above that $M^{(i)}$ \Midef\ are
integrated out.  This is closely related to the fact that the
polynomials $q(z)$ \qtqdef\ are determined (see the discussion
after \anoai).  An ambiguity in $q(z)$ would have appeared as more
fields in the effective superpotential.

\newsec{Matrix model}

We consider the matrix model with the ``action''
 \eqn\matpot{A={\hat N \over \hat g} \left[\Tr W(\hat \Phi) +
 \hat {\tilde Q}_{\tilde f} m^{\tilde f}_f (\hat \Phi) \hat Q^f
 \right] }
where $\hat \Phi$, $\hat {\tilde Q}$ and $\hat Q$ are $\hat
N\times \hat N$, $\hat N\times  N_f$ and $ N_f \times \hat N$
dimensional matrices, and define its free energy $\hat \CF$
through
 \eqn\fedeff{\exp\left(-{\hat N^2\over \hat g^2 }\hat \CF
 \right) =\int d\hat \Phi d \hat Q d\hat {\tilde Q} \exp(-A)}
Eventually, we will take the large $\hat N$ limit with everything
else held fixed.

We will be interested in the resolvents
 \eqn\resolm{\eqalign{
\hat R(z) = & {\hat g \over \hat N} \left\langle \Tr {1
 \over z-\hat \Phi} \right\rangle \cr
 \hat M_{\tilde f}^f(z) = &  \left\langle \hat {\tilde
 Q}_{\tilde f} {1 \over z-\hat \Phi} \hat Q^f\right\rangle
 \cr}}

We perform the following three independent transformations
 \eqn\transf{\eqalign{
 \delta \hat \Phi&= {1\over z-\hat \Phi} \cr
 \delta \hat Q^f &= {1\over z-\hat \Phi} \lambda^f_{f'} \hat
 Q^{f'}\cr
 \delta \hat {\tilde Q}_{\tilde f} &= \hat {\tilde Q}_{\tilde f'}
 \tilde\lambda^{\tilde f'}_{\tilde f}{1\over z-\hat\Phi} \cr}}
to find
 \eqn\anoefm{\eqalign{
 &{\hat N\over \hat g}\left[\Tr {W'(\hat \Phi) \over z-\hat
 \Phi}+\hat {\tilde Q} {m' (\hat \Phi)  \over z-\hat \Phi }\hat Q
 \right] = \left(\Tr {1\over z-\hat \Phi}\right)^2\cr
 & {\hat N\over \hat g }\hat{\tilde Q}{m(\hat \Phi)  \over z-
 \hat \Phi }\lambda \hat Q =\tr \lambda\Tr {1 \over z-\hat \Phi}
 \cr
 & {\hat N\over \hat g } \hat {\tilde Q}\tilde \lambda {m(\hat
 \Phi) \over z- \hat \Phi }\hat Q = \tr \tilde  \lambda\Tr {1
 \over z-\hat \Phi}  \cr
 }}
where we suppressed both the matrix indices and the flavor
indices.  Again we used the convention $A_f^{f'}
B_{f''}^f=(AB)_{f''}^{f'}$, and $\tr$ denotes a trace over the
flavor indices. Taking the expectation values of these equations
and the large $\hat N$ limit they become
 \eqn\anoemf{\eqalign{
 &\left[ W'(z)\hat R(z) \right]_- =\hat R(z)^2\cr
 & \left[ \big(\hat M(z) m(z) \big)_f^{f'}\right]_- = \hat R (z)
 \delta_f^{f'}  \cr
 &\left[\big( m(z)\hat M(z)\big)_{\tilde f}^{\tilde f'} \right]_-=
 \hat R(z) \delta_{\tilde f}^{\tilde f'}
 }}
In the last two equations we used the fact that they are true for
every $\lambda$ and $\tilde \lambda$.

The equations \anoemf\ are the loop equation \migdal\ of the
matrix model.  The first equation describes a closed string
splitting to two closed strings.  The other two equations describe
an open string turning into a closed string.

In solving the first equation for $\hat R(z)$ there is an
ambiguity in a polynomial $\hat f(z)=-4\left[ W'(z)\hat R(z)
\right]_+$.  However, in solving for $\hat M(z)$ the ambiguity in
$\hat q(z)=\left[m(z)\hat M(z) \right]_+$ is fixed by imposing
that $\hat M(z)$ is not singular at the points where $m(z)$ has a
zero eigenvalue.

The equations \anoemf\ are the same as the last three equations
in the gauge theory \anoeff\ and the ambiguity in solving them
(the value of $\hat f(z)$ and the way $\hat q(z)$ is determined)
is also the same.  Therefore we can identify
 \eqn\ident{\eqalign{
 \hat R(z) =& \langle R(z) \rangle \cr
 \hat M(z) =& \langle M (z) \rangle }}

It is interesting that while $\langle\Tr {1\over z- \hat \Phi}
\rangle $ in the matrix model is not identified with $\langle\Tr
{1\over z- \Phi} \rangle $ in the gauge theory but with
$\langle\Tr {W_\alpha^2\over z- \Phi} \rangle $, the matrix model
object $\langle \hat {\tilde Q } {1\over z- \hat \Phi} \hat Q
\rangle $ is identified with its natural gauge theory counterpart
$\langle \tilde Q {1\over z- \Phi}  Q \rangle $.

We now relate the effective superpotential in the gauge theory
\weff\ to the free energy of the matrix model $\hat \CF$.  We
will need the sphere and the disk contributions to the free
energy: $\hat \CF_0=\lim_{\hat N \to \infty}\hat \CF$, $\hat
\CF_1 = \lim_{\hat N \to \infty}{\hat N \over \hat g} (\hat
\CF-\hat \CF_0)$.  $\hat \CF_0$ is computed in the theory without
the fundamental matter.  It was conjectured by Dijkgraaf and Vafa
\DijkgraafDH, and proven in \CDSW\ (see also \DijkgraafXD) that in
this theory $\CF_0=\hat \CF_0$ (recall that $\CF_0$ in the gauge
theory \sphereco\weff\ is the same as in the theory without the
fundamental matter). We now turn to the disk amplitudes.
Following \ident
 \eqn\matcop{\langle \tilde Q_{\tilde f} \Phi^{k-1} Q^f \rangle=
 \langle \hat{\tilde Q}_{\tilde f} \hat \Phi^{k-1} \hat Q^f
 \rangle}
They can be computed in the matrix model and in the gauge theory
as derivatives with respect to the coefficients in $m_f^{\tilde
f}(z)$
 \eqn\cfder{{\partial \CF_1 \over \partial m_{f, k}^{\tilde f}}=
 {\partial \hat \CF_1 \over \partial m_{f, k}^{\tilde f}}}
where we used \phiexp\ and the fact that in the matrix model
$\hat \CF_0$ is independent of $m_{f, k}^{\tilde f}$.  We learn
that $\CF_1=\hat \CF_1+ \delta \CF_1$ with $\delta\CF_1$
independent of $m_{f, k}^{\tilde f}$.  Using the special case
with only $m_{f, k=1}^{\tilde f}= m \delta _f^{\tilde f}$ nonzero
and $m\to \infty$ we easily learn that $\delta\CF_1=0$.

We conclude that
 \eqn\finconf{\eqalign{
 \CF_0=&\hat \CF_0 \cr
 \CF_1=&\hat \CF_1 \cr}}
This establishes the equivalence of the gauge theory and the
matrix model for this case.

\bigskip
\centerline{\bf Acknowledgements}

It is a pleasure to thank D.~Berenstein, F.~Cachazo, M.~Douglas,
D.~Kutasov, G.~Moore and E.~Witten for helpful discussions. This
work was supported in part by DOE grant  \#DE-FG02-90ER40542 to
IAS.

\listrefs

\end